\documentclass[aps,prb,twocolumn,showpacs,preprintnumbers,amsmath,amssymb,floatfix]{revtex4}

\usepackage{graphicx,bm}

\begin{document}

\title{Size effects in the exchange coupling between two electrons in quantum wire quantum dots}

\author{L.-X. Zhang, D. V. Melnikov, S. Agarwal, and J.-P. Leburton}
\affiliation{
Beckman Institute for Advanced Science \& Technology and Department of Electrical and Computer Engineering,
University of Illinois at Urbana-Champaign, Urbana, Illinois 61801
}

\date{\today}
\begin{flushleft}

\end{flushleft}
\begin{abstract}

We theoretically investigate the properties of 
a two-electron system confined  
in the three-dimensional potential of coupled quantum dots formed
in a quantum wire. For this purpose, we implement a variational Heitler-London 
method that minimize the system energies with respect to 
variational parameters in electron trial wavefunctions.
We find that tunneling and exchange couplings exponentially decay 
with increasing inter-dot distance and inter-dot 
barrier height. In the quasi-one-dimensional limit achieved by 
reducing the wire diameter, we find that 
the overlap between the dots decreases, which results in a drop of 
the exchange coupling. We also discuss the validity of our variational 
Heitler-London method with respect to the model potential parameters, and compare
our results with available experimental data to find good agreement between
the two approaches.

\end{abstract}


\maketitle

\section{Introduction}

Very recently, the exploration of new hardware schemes 
at the frontier of solid state quantum computing 
has unveiled a new approach to fabricate coupled quantum 
dots (QDs) with 
controlling gate grid adjacent to an InAs quantum wire (QW).\cite{SamGroup} In 
these device structures, electrons are laterally confined 
({\it i.e.}, perpendicular to the axial direction of 
the wire) by the wire external surfaces (wire diameters are 
tens or even a few nanometers),\cite{Bjork} and longitudinally 
confined in the wire axial direction by the 
electrostatic potential barriers created by the local 
controlling gates. The local 
gate width and separation range from $\sim 10$ 
to $\sim 100$ nm, which results in small effective dot sizes 
and inter-dot separations, so that size quantization 
effects and exchange coupling between the QDs are 
expected to be significantly larger than that in the 
two dimensional electron gas (2DEG) based semiconductor QDs.\cite{Fasth} In quantum wire
quantum dot (QWQD) systems, the distance between the
controlling gates and the QD region ($\sim25$ nm)\cite{SamGroup} is smaller than 
that in 2DEG-based QDs ($\sim100$ nm),\cite{Petta} leading to better electrostatic 
control of the charge (spin) states in the QDs. 
Furthermore, QWQD structures offer linear scalability ({\it i.e.}, with the
linear grid of the controlling gates) instead of the 2D scalability resulting from top or 
side gate patterning in 2DEG-based QDs.\cite{SamGroup}

In laterally coupled 2DEG-based QDs, electron coupling 
occurs between the two QDs in the same 
plane as the 2DEG, and carrier confinement is 
much stronger in the perpendicular direction.\cite{Petta, Elzerman, Hatano} 
In vertically coupled 2DEG-based 
QDs,  carrier confinement is weaker 
in the 2DEG plane than that in the coupling (vertical) 
direction (see {\it e.g.}, Refs. \onlinecite{Pi} and \onlinecite{Bellucci} 
and references therein). 
The electron confinement and coupling defined in 
the fabrication processes of coupled QWQDs 
considerably deviate from those achieved in the 2DEG-based coupled 
QDs: the electrons are strongly confined in the plane 
perpendicular to the axial direction of the wire because of the small wire 
diameter, while quantum mechanical coupling is 
achieved between two quantum wells with relatively weaker confinement, 
because of the controlling gate spacing and biases. 

While a wealth of literature has been 
dedicated to the theoretical study of 2DEG-based 
coupled 
QDs,\cite{Burkard1, Dmm1, Hu, Harju, Szafran, Dybalski, Zhang1, Zhang2, Stopa} 
less attention is paid to 
the QWQD systems. Among all investigated approaches, the 
Heitler-London (HL) technique is relatively simple in its conceptual methodology
to extract the exchange coupling between coupled 
QDs:\cite{Burkard1, Hu} its validity has been 
discussed for systems of various dimensions,\cite{Calderon} 
and efforts have been pursued to 
improve the energy calculation by 
integrating variational parameters in the 
HL method.\cite{Burkard2, Koiller} 

In this paper, we compute
the electronic structure of coupled QWQDs
containing two electrons
with a variational Heitler-London (VHL) method. We 
first construct a three-dimensional (3D) model confinement
potential for the QWQDs and 
introduce three variational parameters in 
the HL wavefunctions that 
account for the specific 3D confinement profile. 
We then numerically minimize the QWQD 
energies with respect to these parameters, 
and obtain the quantum mechanical and
exchange couplings between the two electrons, as well as the
addition energy of the second electron in the dot. In our analysis, special emphasis
is placed on the geometric effects in the coupled QWQDs.
We  discuss
the limitations of our VHL method but indicate its improvement 
over the conventional HL method. We finally compare
our results with the available experimental data. 

\section{model and method}

\begin{figure}[tb]
\includegraphics[width=7cm]{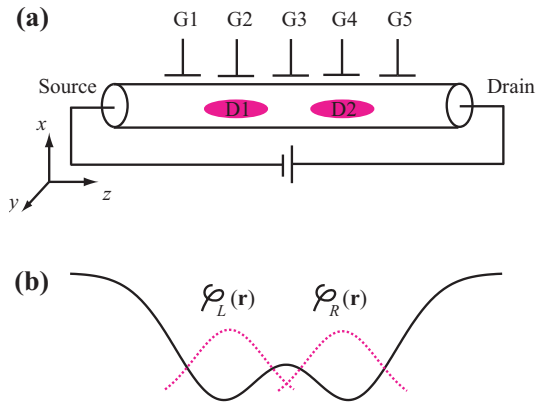}
\caption{\label{fig:fig0} (color online) (a) Schematic of coupled
QDs $D1$ and $D2$ formed in a quantum wire. 
Gates $G1$ and $G5$ define the outer
barriers of the two QDs; $G3$ controls the 
inter-dot coupling; $G2$ and $G4$
are plungers tuning the confinement in each QD. Charging current flows
along the wire from source to drain. (b) Schematic
of the confinement potential of the coupled QDs along the 
$z$ (wire axial) direction. $\varphi_L({\bf r})$ and $\varphi_R({\bf r})$ denote the localized
$s$ states in the left and right QDs, respectively.}
\end{figure}

Figure \ref{fig:fig0}(a) shows a schematic of coupled 
QDs $D1$ and $D2$ formed in a single quantum wire: 
gates $G1$ and $G5$ define the outer
barriers of the QDs, $G3$ controls the inter-dot coupling, 
and $G2$ and $G4$ are used as plunger gates for fine tuning of the 
potential in each QD. The charging current flows from 
source to drain along the wire. The material under 
consideration is InAs, for which we use the electron effective 
mass $m=0.023 m_0$ (Ref. \onlinecite{AEHanson}) and dielectric constant $\epsilon = 14.6$.
Hence, the effective Bohr radius $r_0 = \hbar^2\epsilon/m e^2=33.6$ nm 
and effective Rydberg constant $Ry =m e^4/2\epsilon^2\hbar^2= 1.468$ meV. We assume a parabolic confinement potential in the
$xy$-plane $V({\bm{\rho}}) = {m}\Omega_{\rho}^2{\bm{\rho}}^2 / 2$,
wherein we take $\Omega_{\rho} = \hbar / m (D/2)^2$, and $D$ is the nominal
value of the wire diameter. In the $z$-direction (along which the QDs
are coupled), the confinement potential is modeled by a linear combination of 
three Gaussians:

\begin{eqnarray}
V(z) &=& -V_0\left\{\exp \left[-\frac{(z-d)^2}{l_z^2} \right]+\exp \left[-\frac{(z+d)^2}{l_z^2}\right]\right\}\nonumber\\
&&+V_b \exp\left(-\frac{z^2}{l_{bz}^2}\right),
\label{eqn:V_z}
\end{eqnarray}

\noindent where $V_0$ gives the depth of two Gaussian 
wells describing the confinement of the two individual QDs 
(we fix $V_0=20$ meV), $V_b$ controls the barrier height between the two
wells ($V_b=0$ except otherwise specified), $l_z$ is the radius of each QD, $2d$ is the nominal 
separation between the two QDs,
and $l_{bz}$ denotes the radius of the central barrier.  
A schematic of $V(z)$ is shown in Fig. \ref{fig:fig0}(b) 
by the solid line. The two electrons in the 
coupled QDs are described by the following Hamiltonian:

\begin{equation}
\hat{H} = \hat{H}_{orb}+\hat{H}_Z,
\label{eqn:H_tot}
\end{equation}

\begin{equation}
\hat{H}_{orb}=\hat{h}_1+\hat{h}_2+\frac{e^2}{\epsilon\left|{\bf r_1}-{\bf r_2}\right|},
\label{eqn:H_orb}
\end{equation} 

\begin{equation}
\hat{h}_i= \frac{1}{2m}\left({\bf p_{\rho}}_i+\frac{e}{c}{\bf A}_i\right)^2 + V({\bm \rho}_i)
             +\frac{1}{2m}{{\bf p}_z}_i^2 + V(z_i),
\label{eqn:H_single}
\end{equation}

\begin{equation}
\hat{H}_Z = g\mu_B\sum_{i}{\bf B}\cdot{\bf S}_i.
\label{eqn:H_Z}
\end{equation} 

\noindent Note that we separate the motion of
the electron in the $xy$-plane and in the $z$-direction in the 
single-particle Hamiltonian $\hat{h}_i$. In this work, we only
consider magnetic fields applied in the $z$-direction for 
which ${\bf A} = \left(-yB{\hat x}+xB{\hat y}\right) / 2$. 
Such a magnetic field effectively enhances the confinement 
of the in-plane ($xy$-plane) ground state while preserving 
its cylindrical symmetry. 

In order to obtain the system
energies, we use the following trial wavefunctions:

\begin{equation}
\chi_{\pm}({\bf r}) = \frac{\varphi_L({\bf r})\pm\varphi_R({\bf r})}{\sqrt{2(1 \pm S)}},
\label{eqn:Wv_sp_grnd}
\end{equation}

\begin{equation}
\Psi_{\pm}\left({\bf r}_1, {\bf r}_2\right) = \frac{\varphi_L({\bf r}_1)\varphi_R({\bf r}_2)\pm\varphi_L({\bf r}_2)\varphi_R({\bf r}_1)}{\sqrt{2(1{\pm}S^2)}}.
\label{eqn:WV}
\end{equation}

\noindent In above, $\chi_{+}$, $\chi_{-}$, 
$\Psi_{+}$ and $\Psi_{-}$
denote the single-particle ground and 
first excited states, two-electron singlet and triplet states,
respectively. $S=<\varphi_L|\varphi_R>$ is the overlap between 
$s$ orbitals $\varphi_L({\bf r})$ and $\varphi_R({\bf r})$ localized in
the left and right QDs, respectively, and their specific expressions are

\begin{eqnarray}
\varphi_{L/R}({\bf r})&=&\left(\frac{{m}\omega_\rho}{\pi\hbar}\right)^{\frac{1}{2}}\exp\left[-\frac{{m}\omega_\rho}{2\hbar}(x^2+y^2)\right]\nonumber\\
&&\times\left(\frac{{m}\omega_z}{\pi\hbar}\right)^{\frac{1}{4}}\exp\left[-\frac{{m}\omega_z}{2\hbar}(z \pm a)^2\right].
\label{eqn:WF_1}
\end{eqnarray}

\begin{figure}[tb]
\includegraphics[width=7.8cm]{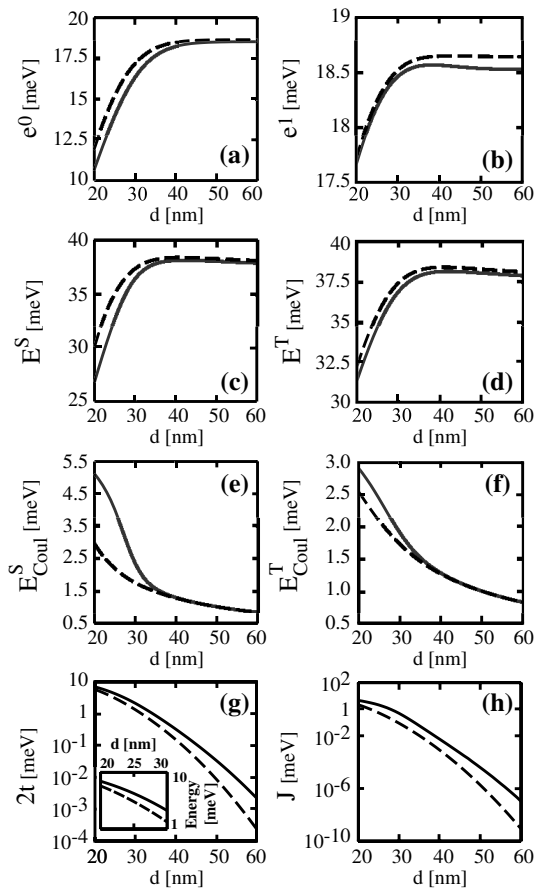}
\caption{\label{fig:fig1} (a) Single-particle ground 
state energy, (b) single-particle first excited state energy, 
(c) two-electron singlet state energy, (d) two-electron 
triplet state energy, (e) Coulomb energy in the singlet state, 
(f) Coulomb energy in 
the triplet state, (g) tunneling coupling $2t$, and 
(h) exchange interaction as a function of the 
half inter-dot separation $d$ for $l_z=30$ nm and 
$D=20$ nm. The inset in (g) shows $2t$ in the zoom-in 
region $20<d<30$ nm. On each panel, 
the solid (dashed) line shows the 
VHL (HL) result.}
\end{figure}

\noindent Figure \ref{fig:fig0}(b) shows
the schematic of $\varphi_L({\bf r})$ and $\varphi_R({\bf r})$ in the $z$-direction
by dashed lines on top
of the potential.
With the variational wavefunctions, we calculate the 
single-particle ground and first excited state energies 
$e^{0/1}=\left<\chi_\pm\left|\hat{h}\right|\chi_\pm\right>$,
two-electron singlet and triplet state energies $E^{S/T}=\left<\Psi_\pm\left|\hat{H}_{orb}\right|\Psi_\pm\right>$.
The detailed expressions of these matrix elements
are given in the Appendix. 

In our VHL approach, 
we use the effective in-plane confinement strength $\omega_\rho$,
$z$-direction confinement strength $\omega_z$ and effective
half inter-dot separation $a$ as variational parameters 
to minimize the system energies.\cite{footnote}
By fixing these variational parameters equal to their nominal values 
$\omega_\rho=\sqrt{\Omega_\rho^2+\omega_c^2}$ 
with $\omega_c=eB/mc$, $\omega_z = \Omega_z=\sqrt{2V_0/ml_z^2}$
and $a=d$, we recover the results from the conventional HL method. 
We calculate the Coulomb energies in the 
singlet and triplet states by

\begin{eqnarray}
E_{Coul}^{S/T} &=& \left<\Psi_{\pm}\left|C\right|\Psi_{\pm}\right>\nonumber\\
&=&\frac{1}{1 \pm S^2}\left(\left<\varphi_L\varphi_R\left|C\right|\varphi_L\varphi_R\right>\right.\nonumber\\
&&\pm\left. \left<\varphi_L\varphi_R\left|C\right|\varphi_R\varphi_L\right>\right),
\label{eqn:Coulomb_all}
\end{eqnarray} 

\noindent where $C = e^2/ \epsilon|{\bf r}_1-{\bf r}_2|$, and we
have used the notation

\begin{eqnarray}
&&\left<\varphi_L\varphi_R\left|C\right|\varphi_L\varphi_R\right>\nonumber\\
&&=\left<\varphi_L({\bf r}_1)\varphi_R({\bf r}_2)\left|C\right|\varphi_L({\bf r}_1)\varphi_R({\bf r}_2)\right>,\\
&&\left<\varphi_L\varphi_R\left|C\right|\varphi_R\varphi_L\right>\nonumber\\
&&=\left<\varphi_L({\bf r}_1)\varphi_R({\bf r}_2)\left|C\right|\varphi_R({\bf r}_1)\varphi_L({\bf r}_2)\right>.
\label{eqn:simple_notation}
\end{eqnarray} 

\noindent The same notation has been used in expressing
the matrix elements in the Appendix. Using both HL and VHL methods, we calculate the
tunnel coupling $2t=e^1-e^0$ and the exchange 
coupling $J=E^T-E^S$. From the two electron wavefunctions,
we compute the electron density as [$\varphi_{L/R}({\bf r})$ are real]

\begin{eqnarray}
\rho^{S/T}({\bf r}_1)&=&2\int\left|\Psi_{\pm}\left({\bf r}_1, {\bf r}_2\right)\right|^2d {\bf r}_2\nonumber\\
&=&\frac{1}{1\pm S^2}\left[\varphi_L^2({\bf r}_1)+\varphi_R^2({\bf r}_1)\right.\nonumber\\
&&\pm\left.2S\varphi_L({\bf r}_1)\varphi_R({\bf r}_1)\right].
\label{eqn:density}
\end{eqnarray}

\section{results}
In Fig. \ref{fig:fig1}, we plot  (a) the single-particle  ground state 
energy $e^0$, (b) single-particle first excited state energy $e^1$,  
(c) two-electron singlet 
state energy $E^S$ and (d) two-electron triplet state energy $E^T$ as a 
function of the half inter-dot separation $d$ for $l_z=30$ nm 
and $D=20$ nm. The solid and dashed lines show the 
results obtained from VHL and HL methods, respectively, from which 
we see that VHL method indeed gives lower system energies than 
the HL method. Here, we note that 
each energy is minimized with respect to a set of its own variational 
parameters. We also note that the single-particle energies are 
positive simply because of the large energy contribution from the in-plane 
confinement: for $D=20$ nm, $\hbar\omega_\rho \approx 33$ meV and 
is changed by less than $1\%$ by varying $d$. 

For $l_z=30$ nm 
and $d=20$ nm, the two Gaussian wells in Eq. (\ref{eqn:V_z}) 
are strongly coupled. As a result, the $z$-direction potential 
has a single minimum at $z=0$, corresponding to a single QD. 
As $d$ increases, a potential barrier between the QDs starts 
to emerge (for $d>21.2$ nm). Meanwhile, the potential minimum 
is raised, and the $z$ confinement in each 
individual QD becomes stronger. The behavior of the 
single-particle energies is a result of these combined effects. 
For example, as $d$ 
increases from $20$ to $38$ nm, both $e^0$ and $e^1$ sharply increase
due to the large increase of the potential minimum [Figs. \ref{fig:fig1}(a) and (b)]. 
For $38<d<60$ nm, $e^0$ still slowly increases, 
while $e^1$ starts to decrease. Our analysis based on 
the variational parameters shows competing effects 
of the kinetic and potential energies in this region: for $e^0$, the 
kinetic energy increase dominates a slight drop of the potential 
energy, whereas for $e^1$, the potential energy increase is 
offset by the drop in the kinetic energy. For very large $d$, both $e^0$ and $e^1$ 
approach a constant value ($18.53$ meV), which corresponds to the 
limit of two decoupled quantum wells. 

\begin{figure}[tb]
\includegraphics[width=7cm]{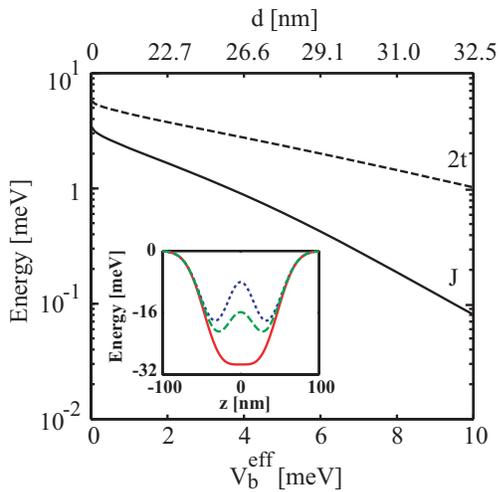}
\caption{\label{fig:fig2} (color online) Main 
panel: Exchange coupling $J$ (solid curve) and 
tunnel coupling $2t$ (dashed curve) as a function 
of the effective barrier height 
$V_b^{eff}$. Values of half inter-dot separation $d$ corresponding 
to different $V_b^{eff}$ values are shown on the upper
horizontal scale. Inset: 
$z$-direction potential profile at $ V_b^{eff}$ values 
$0$ meV (red, solid), $5$ meV (green, dashed) and 
$10$ meV (blue, dotted). Corresponding $V_b$ values 
are $-14.71$ meV, $-1.16$ meV and $6.65$ meV, 
respectively.  Values of other parameter are: 
$D=20$ nm, $d=l_z=30$ nm, and $l_b=30$ nm.}
\end{figure}

The behavior of 
$E^S$ and $E^T$ [Figs. \ref{fig:fig1}(c) and (d)] resembles 
that of $e^0$ and $e^1$, albeit a drop for $d>41$ nm 
is observed for both quantities. The similarity 
implies that the single-particle energies are the dominant 
contributions to  $E^S$ and $E^T$, 
whereas the decrease of Coulomb energy with 
increasing $d$ [Figs. \ref{fig:fig1} (e) and (f)] has a 
minor influence. It is seen that at 
fixed $d$, the Coulomb interaction is stronger in the singlet 
state, due to the larger overlap ($S$) in the two-electron wavefunction, 
which is a signature of the Pauli exclusion principle. 

In Figs. \ref{fig:fig1} (g) and (h), we plot the tunnel coupling 
$2t$ and exchange coupling $J$ as a function 
of $d$, respectively, both of which exhibit exponential decay with increasing $d$ 
(strictly speaking, the decay is slightly slower than exponential). 
In these figures, the solid (dashed) line  
corresponds to the VHL (HL) result. A much larger decrease of $J$ 
($\sim 10^{-8}$)  than $2t$ ($\sim 10^{-4}$) as
$d$ increases from $20$ to $60$ nm agrees qualitatively with the Hubbard
model $J \propto (2t)^2/U_H$, 
assuming that the intra-dot Coulomb
interaction $U_H$ retains the same order of magnitude as
$d$ varies. Figs. \ref{fig:fig1} (g) and (h) show a large difference between 
the tunnel and exchange couplings obtained by using
the HL and VHL methods,  from which we notice that the HL method  
substantially {\it underestimates} 
the coupling between the two electrons,\cite{Sousa} especially for large
inter-dot separations. For example, at $d=60$ nm, the VHL result 
of $2t$ ($J$) is $\sim10$ ($\sim 100$) times of the HL result.

The inset in Fig. \ref{fig:fig2} indicates that both the effective 
barrier height $V_b^{eff}$ ({\it i.e.}, the energy difference between the
minima of the potential and its value at $z=0$) 
and the distance between the two QDs ({\it i.e.,} the distance 
between the two minima of the potential) become larger as 
$V_b$ is increased. Consequently, both 
$2t$ and $J$ exhibit nearly exponential decay\cite{Hu} with 
increasing $V_b^{eff}$ as shown in the main panel 
of Fig. \ref{fig:fig2}, similar to the quasi-exponential drop of these 
two quantities with increasing 
QD separation $2d$ [cf. Figs. \ref{fig:fig1}(g) and (h)]. Again, we observe 
that $J$ decays at a much faster rate than $2t$. In experimental 
QWQD devices, the effective barrier height between the two QDs 
can be tuned by varying the central gate bias,\cite{SamGroup} 
and our analysis shows that the magnitude of the exchange 
coupling can be controlled by proper biasing the central gate
as in 2DEG-based coupled QDs.\cite{Hu}

\begin{figure}[tb]
\includegraphics[width=7cm]{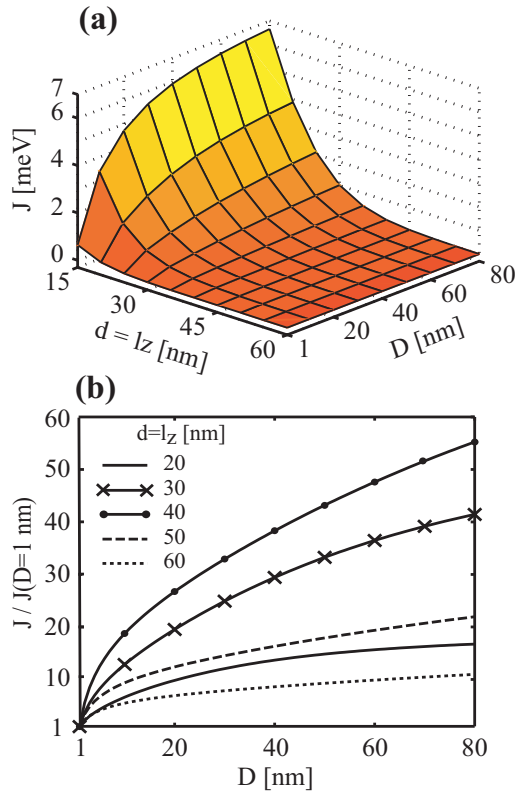}
\caption{\label{fig:fig3} (color online) (a) 
Exchange coupling $J$ as a function of wire 
diameter $D$ and half separation between the QDs 
$d$, which is set equal to QD radius $l_z$ 
($d=l_z$). (b) $J$ as a function of $D$ for different $d=l_z$ values 
(shown in the figure). The $J$ value on each 
curve is normalized to its value at $D=1$ nm. 
For $(d=l_z)=20$, $30$, $40$, $50$, $60$ nm, 
$J(D=1 \text{nm})=2.33 \times 10^{-1}$, $2.47 \times 10^{-2}$, 
$3.53 \times 10^{-3}$, $1.37 \times 10^{-3}$, 
$4.81 \times 10^{-4}$ meV, respectively.}
\end{figure}

\begin{figure}[tb]
\includegraphics[width=7cm]{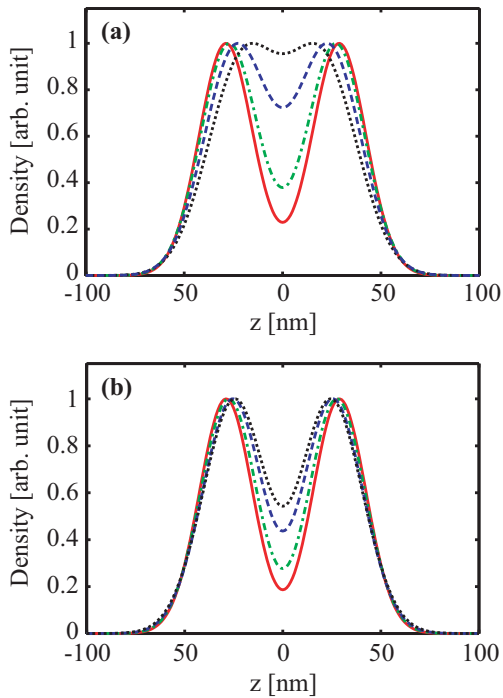}
\caption{\label{fig:fig4} (color online) Electron 
density plot in the $z$-direction for (a) singlet 
and (b) triplet states at $d=l_z=30$ nm. In each 
figure, the density is plotted at $D=1$ nm (red, solid), 
$D=10$ nm (green, dashed-dotted), $D=40$ nm (blue, 
dashed) and $D=80$ nm (black, dotted). For each $D$, 
the density is normalized to its peak value.}
\end{figure}

Figure \ref{fig:fig3}(a) displays the exchange coupling $J$ as 
a function of both the wire diameter 
$D$ and the half separation ($d$) between the two QDs. 
Here, we set $d=l_z$ noting that in experiments 
coupled QWQDs are defined on top of a linear gate 
grid with a particular periodicity,\cite{SamGroup} which indicates that the 
effective QD size and inter-dot separation are approximately 
the same. For the confinement potential given by 
Eq. (\ref{eqn:V_z}), this configuration 
leads to a constant effective barrier height of $5.68$ meV, 
independent of the value of $d=l_z$. The nominal confinement 
strength for a single Gaussian well ($V_0=20$ meV) with 
$l_z=15$ nm ($0.45 r_0$) and $60$ nm ($1.78 r_0$) is 
$\hbar \Omega_z=24.27$ and $6.07$ meV, respectively. 
For a wire diameter $D=1$ nm, the nominal confinement is 
$\hbar\Omega_\rho=1.33\times10^4$ meV, which physically 
corresponds to the {\it quasi-1D limit} of the systems 
with aspect ratio $(\lambda_\rho / \lambda_z = \sqrt{\Omega_z / \Omega_\rho}) <0.05$ for the 
investigated range of $d=l_z$ from $15$ to $60$ nm. In the opposite limit, where 
$D=80$ nm, $\hbar\Omega_\rho=2.07$ meV, the aspect ratio 
$\lambda_\rho / \lambda_z >1.71$.  At fixed $D$, $J$ exhibits exponential 
decay with $d=l_z$ in Fig. \ref{fig:fig3}(a), where it is also observed that 
$J$ decreases with decreasing $D$ at fixed $d=l_z$. This trend is shown 
explicitly in Fig. \ref{fig:fig3}(b) for different $d=l_z$. For comparison, the 
data on each curve are normalized to the value of $J$ at 
$D=1$ nm. At fixed $d=l_z$, as $D$ is decreased from 
$80$ nm, $J$ decreases, and the decreasing rate becomes larger
as $D$ approaches $1$ nm, which 
is the $quasi-1D$ limit. The faster dropping rate of $J$ 
near $D=1$ nm is due to $\Omega_\rho \propto 1/D^2$, 
and the influence of the variation of $\Omega_\rho$ on $J$ becomes 
stronger at smaller $D$ (through the Coulomb interaction). 
Here, we note that although 
the general trend of $J$ is to decrease as $D$ is made smaller, the decreasing 
rates are much larger for intermediate $d=l_z$ values 
than for small or large values.

\begin{figure}[tb]
\includegraphics[width=7.5cm]{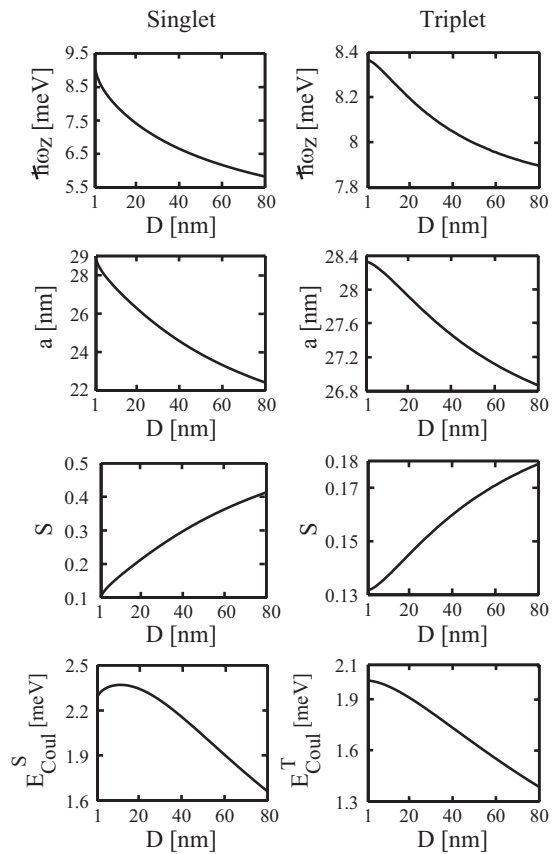}
\caption{\label{fig:fig5} Variational parameters 
$\omega_z$ (shown as $\hbar\omega_z$), half separation 
$a$, the overlap $S$, and the Coulomb energies 
as a function of $D$ at $d=l_z=30$ nm.  
Left (right) panels are for the singlet (triplet) state.}
\end{figure}

These effects of the wire diameter variation on the 
exchange coupling are rather unexpected as they show  
that $J$ depends on the wire confinement 
perpendicular to the coupling direction. In fact, we find 
that the $D$ variation not only changes $\omega_\rho$, 
but also induces significant 
changes in  $\omega_z$ and 
$a$, which minimize the singlet and triplet state energies. 
One can directly visualize such changes by 
inspecting the electron density variation with respect 
to the wire diameter. In Fig. \ref{fig:fig4}, we plot 
the electron 
density [Eq. (\ref{eqn:density})] for different $D$ 
values ($d=l_z=30$ nm) in (a) the singlet and (b) triplet states, 
respectively. For the singlet state, as 
$D$ decreases, the separation between the two 
density peaks becomes larger, and the width 
of each peak becomes smaller. Consequently, the overlap 
between the two electrons is reduced. Similar effects 
are observed in the density of the triplet state to a less extent. 

In Fig. \ref{fig:fig5}, we plot the $D$ dependence 
of $\omega_z$ and $a$ on  the top two rows. 
Both variational parameters increase as $D$ is 
reduced, and the relative increase is more significant
in the singlet state than the triplet state. As a 
consequence, the overlap 
$S=<\varphi_L|\varphi_R>=\exp(-m \omega_z a^2/ \hbar)$ between 
the localized $s$ states 
decreases with decreasing $D$ in both states, and the relative decrease
is larger in the singlet state [Fig. \ref{fig:fig5}, third row]. 
Despite this effect, the 
Coulomb interaction ($E_{Coul}$) becomes 
stronger with decreasing $D$ [Fig. \ref{fig:fig5}, bottom row] for 
both states, which is due to 
the reduced size in the $xy$-plane. We 
also performed analysis for different $d=l_z$ and 
observed similar behavior as shown in 
Figs. \ref{fig:fig4} and \ref{fig:fig5}.

\begin{figure}[tb]
\includegraphics[width=7cm]{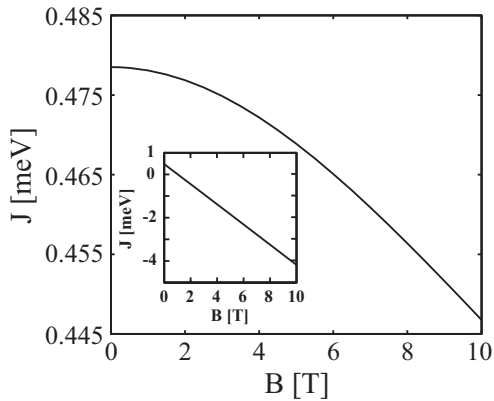}
\caption{\label{fig:fig6} Main panel: Exchange coupling 
as a function of the magnetic field applied along the 
wire without the Zeeman effect for 
$D=20$ nm, $d=l_z=30$ nm. Inset: same as main panel but 
with Zeeman effect.} 
\end{figure}

In general, the influence of the $D$ variation 
on the exchange energy results from the fact that the 
two electrons in the 3D QWQD system respond to 
the variation of a single external parameter by 
adapting all the variational parameters via the 
minimization of the system energy. The response varies 
depending upon the values of other fixed external parameters, 
which leads to the different decreasing 
rates observed in Fig. \ref{fig:fig3}(b), for instance.

The in-plane electron confinement can also be 
enhanced by applying a magnetic field ($B$) along 
the wire without reducing the wire diameter. 
As with reducing $D$, $J$ drops 
with increasing $B$ as seen in Fig. \ref{fig:fig6}, main panel. 
The drop is nearly linear at large $B$, which is  
smaller than the drop rate  when 
$D$ approaches $1$ nm [cf. Fig. \ref{fig:fig3}(b)]. 
This is because the in-plane effective (variational) confinement 
strength 
$\omega_\rho \approx \sqrt{\Omega_\rho^2+\omega_c^2}$ 
and $\omega_c \propto B$, while $\Omega_\rho \propto 1/D^2$. 
It should be pointed out that the relatively 
small $J$ drop in Fig. \ref{fig:fig6} 
is obtained in the absence of the Zeeman effect, and 
it is well known that unlike the small $g$ 
factor in GaAs ($g \approx -0.44$), InAs QWQD has 
a much larger $g$ factor ($2$ to $15.5$),\cite{Bjork} 
for which the Zeeman effect
is dominant over the orbital effect in the 
$J$ dependence on $B$. For example, the inset of Fig. \ref{fig:fig6}
shows that for $g=8$,\cite{gfactor}
the Zeeman effect totally smears out the orbital effect 
illustrated in the main panel of Fig. \ref{fig:fig6}, which leads to a negative $J$ 
for $B>1.1$ T.

Because we model the confinement in the $xy$-plane
by a two-dimensional harmonic oscillator potential, 
the single-particle levels in that plane are given by the Fock-Darwin 
spectrum, whereby the energy separation between the ground 
and first excited states
decreases as $B$ increases (in contrast, this separation
increases with decreasing $D$). 
In our calculations in Fig. \ref{fig:fig6}, 
we take $D=20$ nm and $d=l_z=30$ nm.
At $B=10$ T, the
separation is $16.44$ meV, which is considerably larger 
than the sum of the single-particle energy separation in the 
$z$-direction ($2.11$ meV) and the Coulomb energy 
in the triplet state ($1.91$ meV). This observation 
validates the assumptions of the HL 
method in which the wavefunctions are taken as linear 
combination of localized Gaussians separated in the $z$-direction, and
only the ground state in the $xy$-plane is taken into account. 

\begin{figure}[tb]
\includegraphics[width=7cm]{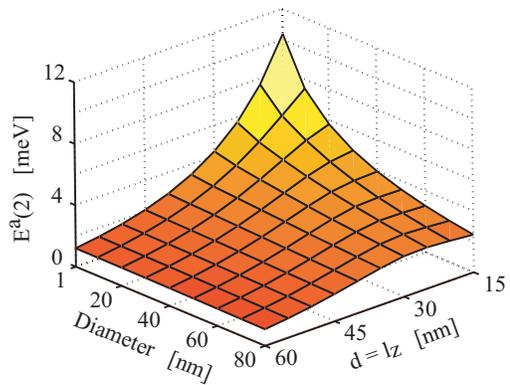}
\caption{\label{fig:fig7} (color online) Addition energy of the 
second electron $E^a(2)$ as a fucntion of wire diameter $D$
and half inter-dot separation $d=l_z$.}
\end{figure}

Experimentally, the measurement of the addition 
energy is frequently performed to probe the energy levels 
of the QD.\cite{Fasth} The addition energy of the $N$-th 
electron is defined as $E^a(N)=\mu(N)-\mu(N-1)$, 
where $\mu(N)$ is the chemical potential of an $N$-electron 
QD. Within the VHL 
method, we are able to calculate the addition energy 
of the second electron as 
$E^a(2)=\mu(2)-\mu(1)=E^S-2e^0$, 
where $E^S$ and $e^0$ denote the singlet 
state energy and the single-particle ground 
state energy, respectively. We plot $E^a(2)$ 
as a function of the geometric parameters $d=l_z$ 
and $D$ in Fig. \ref{fig:fig7}. In general, as 
the QDs become larger in size (larger $d=l_z$ or $D$), 
the addition energy decreases, for both Coulomb 
interaction and size quantization effects are reduced. 
We find (not shown) that at fixed $d=l_z$ and $D$, the Coulomb 
energy between the two electrons are uniformly 
smaller than $E^a(2)$, which is due to the size quantization 
effects in the coupled QWQDs. 

\section{Discussions}
\label{sec:discussion}
\subsection{Limitation of the variational Heitler-London method}
\label{sec:discussion_a}

As an inherent drawback of the HL 
method, our variational scheme breaks down when 
the overlap between the localized $s$ states is large, which occurs
for small inter-dot separations. For example, in our
calculations of the system energies, the VHL method fails
for $(d=l_z)<12$ nm ($0.57r_0$) independent of $D$.
A signature of the VHL approach breakdown at small $d$ is that
the variational parameter $a$ becomes zero in the minimization 
process. This numerical behavior stems from
the fact that at small $d$ a global minimum in the system energies 
does not exist for the physical range of $a$, given the 
expression of the variational wavefunction. 
We note that this shortcoming in the HL 
method in not apparent in the conventional HL approach.
As long as $(d=l_z)>0$, one can still use 
the HL method (without variation) to calculate the system energies 
even though the obtained result is likely to be unphysical. 

In Ref.~\onlinecite{Calderon}, it was pointed out 
that the HL method breaks down 
as the quantity 
$c=\sqrt{\pi/2}(e^2/\epsilon a_B) / \hbar\omega_0$ 
($a_B = \sqrt{\hbar/m\omega_0}$) is larger 
than $1.95$, $2.8$, and $5.8$ for coupled QDs 
with harmonic oscillator confinement 
$\hbar\omega_0$ in each direction for 1D, 
2D and 3D potential models, respectively (this is 
an extension of the result in Ref. \onlinecite{Burkard1}). 
We investigate $l_z$ from $15$ to $60$ nm, which corresponds 
to $c$ ranging from $0.44$ to $0.87$, and is uniformly
smaller than the smallest breakdown value $c=1.95$. However, 
as a check of this criterion, we extend our 
calculation to very large value of $d=l_z$ 
and find that for $D=20$ nm, $J$ becomes very
 noisy and oscillates randomly for 
$(d=l_z)>206$ nm, for which the 
variational parameter $\hbar\omega_z$ is 
$1.553$ meV, corresponding to $c=1.723$, 
which is similar to the 1D limit claimed above. 
However, at this point, $J \sim 10^{-14}$ meV, which bears no 
practical interest.

\subsection{Comparison with experiments}

In recent experiments on InAs QWQDs, 
$J=2.8$ to $3.2$ meV was reported for a single 
QD formed in a wire with effective 
harmonic confinement strength $\hbar\Omega_z=6.3$ meV 
(corresponding to confinement length 
$2\lambda_z=2\sqrt{\hbar / m \Omega_z}=46$ nm) 
and  $\hbar\Omega_\rho=40$ meV 
($2\lambda_\rho=2\sqrt{\hbar / m \Omega_\rho}=18$ nm).\cite{Fasth} 
By fitting these values in our model 
($D=18$ nm, $V_0=41.6$ meV, $V_b=0$ meV, 
$d=0$ nm and $l_z=117.9$ nm), we obtain 
$J=3.51$ meV, which is comparable to the 
experimental result. 

We note that $J \sim 3$ meV as obtained above is the result for a single QD
with potential minimum at $z=0$.\cite{footnote2} 
For double QDs with $D=20$ and $d=l_z=30$ nm, we obtain $J \sim 0.5$ 
meV (Fig. \ref{fig:fig6}), which corresponds to a time scale 
$(\tau_J = \hbar / J) \sim 1.3$ ps, on the same order as the
reported spin decoherence time $T_2=0.5-1$ 
ps in InAs QWQDs\cite{AEHanson} and much smaller than 
the reported spin dephasing time $T_2^*=50-500$ 
ps  in self-assembled InAs 
QDs.\cite{MBGroup} 

\section{Conclusion}

By introducing variational parameters in 
the HL trial wavefunctions, we achieved lower energies 
of coupled QWQD 
system than those calculated by conventional HL method with 
the relative difference in the tunnel and exchange 
couplings exceeding $100\%$.  
As in coupled GaAs QDs based on 2DEG, tunnel
and exchange couplings 
exhibit exponential decay with increasing inter-dot distance 
or barrier height. Due to the 3D nature of the 
system, increasing the confinement in the in-plane directions 
reduces the overlap of the two electrons in the coupling 
direction (along the wire), which results in the decrease of the 
exchange coupling. For QDs with different sizes, the addition 
energy of the second electron is found to be uniformly larger
than the two-electron Coulomb interaction because of size quantization
effects. 
By fitting the model potential to 
experimental parameters, we obtain exchange coupling in agreement
with experimental data. Experimental structures based on InAs 
QWQDs may benefit from the relatively 
large exchange coupling towards quantum computing 
applications.

\begin{acknowledgments}
This work is supported by the DARPA QUIST program through 
ARO Grant DAAD 19-01-1-0659. The authors thank the 
Material Computational Center at the University of Illinois 
through NSF Grant DMR 99-76550. LXZ thanks the University
of Illinois Research Council, the Beckman
Institute and the Computer Science and Engineering program at
the University of Illinois. 
\end{acknowledgments}

\appendix*
\section{}
\label{sec:appendix}

The single particle Hamiltonian $\hat{h}$ can be rewritten as

\begin{equation}
\hat{h} = \hat{h}_{L/R}^0 + W_{L/R},
\end{equation}
\begin{eqnarray}
\hat{h}_{L/R}^0&=&\frac{1}{2m}\left({\bf p_{\rho}}+\frac{e}{c}{\bf A}\right)^2 + \frac{1}{2}{m}\omega_{\rho 0}^2{\bm \rho}^2\nonumber\\ 
&&+ \frac{1}{2m}p_{z}^2 + \frac{1}{2}{m}\omega_{z}^2(z \pm a)^2,
\label{P_H}
\end{eqnarray}
\begin{equation}
W_{L/R} = \frac{1}{2}{m}\left(\Omega_{\rho}^2- \omega_{\rho 0}^2\right){\bm \rho}^2 + V(z)-\frac{1}{2}{m}\omega_{z}^2(z \pm a)^2.
\label{W_single}
\end{equation}

\noindent Since $\hat{h}_{L/R}^0\varphi_{L/R}({\bf r})=E_0\varphi_{L/R}({\bf r})$ and  
$E_0 = \hbar\omega_{\rho}+\hbar\omega_z / 2$, we only need to calculate the matrix
element of $W_{L/R}$. Thus, we have

\begin{widetext}
\begin{eqnarray}
e^{0/1} &=& \left<\chi_\pm\left|\hat{h}\right|\chi_\pm\right>\nonumber\\
&=&\frac{1}{2}\hbar\omega_\rho+\frac{1}{4}\hbar\omega_z+\frac{\hbar}{2\omega_\rho}\left(\Omega_\rho^2+\frac{\omega_c^2}{4}\right)+\frac{1}{1\pm S}\left\{-V_0\left(\frac{\hbar}{m\omega_z l_z^2}+1\right)^{-\frac{1}{2}}\exp\left[-\left(\frac{1}{l_z^2}+\frac{m\omega_z}{\hbar}\right)^{-1}\frac{m\omega_z}{\hbar l_z^2}(a+d)^2\right]\right.\nonumber\\
&&-V_0\left(\frac{\hbar}{m\omega_z l_z^2}+1\right)^{-\frac{1}{2}}\exp\left[-\left(\frac{1}{l_z^2}+\frac{m\omega_z}{\hbar}\right)^{-1}\frac{m\omega_z}{\hbar l_z^2}(a-d)^2\right]\nonumber\\
&&+V_b\left(\frac{\hbar}{m\omega_z l_{bz}^2}+1\right)^{-\frac{1}{2}}\exp\left[-\left(\frac{1}{l_{bz}^2}+\frac{m\omega_z}{\hbar}\right)^{-1}\frac{m\omega_z}{\hbar l_{bz}^2}a^2\right]\mp 2SV_0\left(\frac{\hbar}{m\omega_z l_z^2}+1\right)^{-\frac{1}{2}}\exp\left[-\left(\frac{1}{l_z^2}+\frac{m\omega_z}{\hbar}\right)^{-1}\frac{m\omega_z}{\hbar l_z^2}d^2\right]\nonumber\\ 
&&\pm \left. SV_b\left(\frac{\hbar}{m\omega_z l_{bz}^2}+1\right)^{-\frac{1}{2}}\mp\frac{S}{2}m\omega_z^2a^2\right\}.
\label{eqn:E_sp}
\end{eqnarray}
\end{widetext}

\noindent In above, $S  =  \int d{\bf r}\varphi_{L}^*({\bf r})\varphi_{R}({\bf r}) = \exp\left(-m\omega_za^2/\hbar\right)$ is 
the overlap between the two localized $s$ states.

The singlet and triplet energies are evaluated in a similar fashion and the results are:
\begin{widetext}
\begin{eqnarray}
E^{S/T} &=& \left<\Psi_{\pm}\left|\hat{H}_{orb}\right|\Psi_{\pm}\right>\nonumber\\
&=&2E_0+ \frac{1}{1\pm S^2}\left[\left<\varphi_{L}\varphi_{R}|W_1+W_2+W_3+C|\varphi_{L}\varphi_{R}\right>\right.\pm \left. Re\left<\varphi_{L}\varphi_{R}|W_1+W_2+W_3+C|\varphi_{R}\varphi_{L}\right>\right],
\label{E_two}
\end{eqnarray}
\end{widetext}
\begin{widetext}
\begin{eqnarray}
\left<\varphi_{L}\varphi_{R}\left|W_1\right|\varphi_{L}\varphi_{R}\right> &=& \left<\varphi_{L}\varphi_{R}\left|V(z_1)+V(z_2)\right|\varphi_{L}\varphi_{R}\right>\nonumber\\
&=&-2V_0\left(\frac{\hbar}{m\omega_z l_z^2}+1\right)^{-\frac{1}{2}}\left\{ \exp\left[-\left(\frac{1}{l_z^2}+\frac{m\omega_z}{\hbar}\right)^{-1}\frac{m\omega_z}{\hbar l_z^2}(a+d)^2\right]\right.\nonumber\\
&&+\left. \exp\left[-(\frac{1}{l_z^2}+\frac{m\omega_z}{\hbar})^{-1}\frac{m\omega_z}{\hbar l_z^2}(a-d)^2\right]\right\}\nonumber\\
&&+2V_b\left(\frac{\hbar}{m\omega_z l_{bz}^2}+1\right)^{-\frac{1}{2}}\exp\left[-\left(\frac{1}{l_{bz}^2}+\frac{m\omega_z}{\hbar}\right)^{-1}\frac{m\omega_z}{\hbar l_{bz}^2}a^2\right],\\
\left<\varphi_{L}\varphi_{R}\left|W_1\right|\varphi_{R}\varphi_{L}\right> &=& \left<\varphi_{L}\varphi_{R}\left|V(z_1)+V(z_2)\right|\varphi_{R}\varphi_{L}\right>\nonumber\\
&=&2\exp\left(-2\frac{m\omega_z}{\hbar}a^2\right)\left\{-2V_0\left(\frac{\hbar}{m\omega_z l_z^2}+1\right)^{-\frac{1}{2}}\exp\left[-\left(\frac{1}{l_z^2}+\frac{m\omega_z}{\hbar}\right)^{-1}\frac{m\omega_z}{\hbar l_z^2}d^2\right]\right.\nonumber\\
&&+\left. V_b\left[\frac{\hbar}{m\omega_z l_{bz}^2}+1\right]^{-\frac{1}{2}} \right\},
\end{eqnarray}
\end{widetext}
\begin{widetext}
\begin{eqnarray}
\left<\varphi_{L}\varphi_{R}\left|W_2\right|\varphi_{L}\varphi_{R}\right>&=&\left<\varphi_{L}\varphi_{R}\left|-\frac{1}{2}{m}\omega_{z}^2(z_1 + a)^2-\frac{1}{2}{m}\omega_{z}^2(z_2 - a)^2\right|\varphi_{L}\varphi_{R}\right>\nonumber\\
&=&-\frac{1}{2}\hbar\omega_z,\\
\left<\varphi_{L}\varphi_{R}\left|W_2\right|\varphi_{R}\varphi_{L}\right>&=&\left<\varphi_{L}\varphi_{R}\left|-\frac{1}{2}{m}\omega_{z}^2(z_1 + a)^2-\frac{1}{2}{m}\omega_{z}^2(z_2 - a)^2\right|\varphi_{R}\varphi_{L}\right>\nonumber\\
&=&-m\omega_z^2\left(\frac{\hbar}{2m\omega_z}+a^2\right)\exp\left(-2\frac{m\omega_z}{\hbar}a^2\right),\\
\left<\varphi_{L}\varphi_{R}\left|W_3\right|\varphi_{L}\varphi_{R}\right>&=&\left<\varphi_{L}\varphi_{R}\left|\frac{1}{2}{m}\left(\Omega_{\rho}^2- \omega_{\rho 0}^2\right)\left({\bm \rho}_1^2+{\bm \rho}_2^2\right)\right|\varphi_{L}\varphi_{R}\right>\nonumber\\
&&=\frac{\hbar}{\omega_\rho}\left(\Omega_\rho^2+\frac{\omega_c^2}{4}-\omega_\rho^2\right),\\
\left<\varphi_{L}\varphi_{R}\left|W_3\right|\varphi_{R}\varphi_{L}\right>&=&\left<\varphi_{L}\varphi_{R}\left|\frac{1}{2}{m}\left(\Omega_{\rho}^2- \omega_{\rho 0}^2\right)\left({\bm \rho}_1^2+{\bm \rho}_2^2\right)\right|\varphi_{R}\varphi_{L}\right>\nonumber\\
&=&\frac{\hbar}{\omega_\rho}\left(\Omega_\rho^2+\frac{\omega_c^2}{4}-\omega_\rho^2\right)\exp\left(-2\frac{m\omega_z}{\hbar}a^2\right).
\label{eqn:W123}
\end{eqnarray}
\end{widetext}

The Coulomb matrix elements in Eq. (\ref{eqn:Coulomb_all}) are given by

\begin{widetext}
\begin{eqnarray}
\left<\varphi_{L}\varphi_{R}\left|C\right|\varphi_{L}\varphi_{R}\right>&=&\left<\varphi_{L}\varphi_{R}\left|\frac{e^2}{\epsilon\left|{\bf r_1}-{\bf r_2}\right|}\right|\varphi_{L}\varphi_{R}\right>\nonumber\\
&&=\frac{e^2}{\epsilon}\left(\frac{2m\omega_z}{\pi\hbar}\right)^{\frac{1}{2}}\int_0^1\frac{1}{1+\left(\frac{\omega_z}{\omega_\rho}-1\right)t^2}\exp\left(-2\frac{m\omega_z}{\hbar}a^2t^2\right)dt,\\
%\end{eqnarray}
%\begin{eqnarray}
\left<\varphi_{L}\varphi_{R}\left|C\right|\varphi_{R}\varphi_{L}\right>&=&\left<\varphi_{L}\varphi_{R}\left|\frac{e^2}{\epsilon\left|{\bf r_1}-{\bf r_2}\right|}\right|\varphi_{R}\varphi_{L}\right>\nonumber\\
&&= \frac{e^2}{\epsilon}\left(\frac{2m\omega_z}{\pi\hbar}\right)^{\frac{1}{2}}\exp\left(-2\frac{m\omega_z}{\hbar}a^2\right)\int_0^1 \frac{1}{1+\left(\frac{\omega_z}{\omega_\rho}-1\right)t^2}dt\nonumber\\
&&=\frac{e^2}{\epsilon}\left(\frac{2m\omega_z}{\pi\hbar}\right)^{\frac{1}{2}} \exp\left(-2\frac{m\omega_z}{\hbar}a^2\right) \times \left\{\begin{array}{lcl}{\left(1-\frac{\omega_z}{\omega_\rho}\right)^{-\frac{1}{2}} \text{arctanh} \left[\left(1-\frac{\omega_z}{\omega_\rho}\right)^{\frac{1}{2}}\right]}
& \mbox {for} & \omega_z<\omega_\rho \\\\ {\left(\frac{\omega_z}{\omega_\rho}-1\right)^{-\frac{1}{2}} \arctan\left[\left(\frac{\omega_z}{\omega_\rho}-1\right)^{\frac{1}{2}}\right]} & \mbox {for} & \omega_z>\omega_\rho\\\\
1 & \mbox {for} & \omega_z=\omega_\rho. \end{array}\right.\nonumber\\
\label{eqn:Coulomb_D_2}
\end{eqnarray}
\end{widetext}

\noindent Practically, the one dimensional 
integrals in Eqs. (A.12) and (\ref{eqn:Coulomb_D_2})] are numerically evaluated using
adaptive quadratures. We note that in the 1D limit ($\omega_\rho \rightarrow \infty$),
the integrals have logarithmic divergence,\cite{Calderon} while they both approach zero in the 
oppsite limit ($\omega_\rho \rightarrow 0$). For $\omega_\rho=\omega_z=\omega_0$,
the integrals in Eqs. (A.12) and (\ref{eqn:Coulomb_D_2}) simplify to 
$[e^2/(2a\epsilon)] \text{Erf}(a\sqrt{2m\omega_0/\hbar})$ and 
$(e^2/ \epsilon ) \sqrt{2m\omega_0 / (\pi \hbar)})\exp(-2m\omega_0 a^2 / \hbar)$, respectively.
These results are identical to the results in 
Ref. \onlinecite{Calderon}, where the Coulomb matrix elements 
were calculated between coupled spherically symmetric
Gaussian trial wavefunctions.


\begin{thebibliography}{99}
\bibitem{footnote} 
An intial
guess for the variational parameters is given near their nominal values
and then we use the \texttt{fminsearch} subroutine in MATLAB which
employs the Nelder-Mead simplex method to search
for the minima of the system energies with respect to these parameters. It turns out that unless the overlap
between the localized $s$ states is too large, 
the simplex method always finds a global minimum 
within the physically reasonable range of the variational 
parameters.
\bibitem{gfactor} We take
$g=8$ measured in QDs with similar geometry,\cite{Fasth} 
although the linear dependence of $J$ on magnetic field
is clear even for $g=2$.
\bibitem{footnote2} Recently, the VHL
method has been applied to calculate the 
two-electron energies in a single two-dimensional elliptical
QD. By comparing the results to numerical exact diagonalization results,
it is shown that the VHL method is successful in reproducing 
the magnetic field dependence of $J$.\cite{Agarwal}
\end{thebibliography}
\end{document}